\documentclass[conference]{IEEEtran}

\usepackage{amsmath,amssymb,amsfonts}
\usepackage{algorithmic}
\usepackage{algorithm}
\usepackage[font=footnotesize,labelfont=rm,textfont=rm]{subfig}
\usepackage{textcomp}

\usepackage{graphicx}
\usepackage{cite}
\usepackage{balance}
\usepackage{multirow}
\usepackage{makecell}
\usepackage{xcolor}

\usepackage{fancyhdr}

\fancypagestyle{firstpage}{
  \fancyhf{} 

  \fancyhead[C]{This paper has been accepted by IEEE GLOBECOM 2025}
  
}

\def\BibTeX{{\rm B\kern-.05em{\sc i\kern-.025em b}\kern-.08em
		T\kern-.1667em\lower.7ex\hbox{E}\kern-.125emX}}

\begin{document}
\title{A Real-time Data Collection Approach for 6G AI-native Networks}

\setlength{\textfloatsep}{8pt}
\setlength{\intextsep}{2pt}

\author{
	\IEEEauthorblockN{
		Shiwen He\textsuperscript{1,2},
		Haolei Dong\textsuperscript{1},
        Liangpeng Wang\textsuperscript{2},
        Zhenyu An\textsuperscript{2}
		}
	\IEEEauthorblockA{
        \textsuperscript{1}School of Computer Science and Engineering, Central South University, Changsha, China, 410083
        \\\textsuperscript{2}Purple Mountain Laboratories, Nanjing, China, 211100
        \\Email: shiwen.he.hn@aliyun.com, dong.hl@csu.edu.cn, \{wangliangpeng, anzhenyu\}@pmlabs.com.cn}

}

\maketitle

\thispagestyle{firstpage}

\begin{abstract}
During the development of the Sixth Generation (6G) networks, the integration of Artificial Intelligence (AI) into network systems has become a focal point, leading to the concept of AI-native networks. High-quality data is essential for developing such networks. Although some studies have explored data collection and analysis in 6G networks, significant challenges remain, particularly in real-time data acquisition and processing. This paper proposes a comprehensive data collection method that operates in parallel with bitstream processing for wireless communication networks. By deploying data probes, the system captures real-time network and system status data in software-defined wireless communication networks. Furthermore, a data support system is implemented to integrate heterogeneous data and provide automatic support for AI model training and decision making. Finally, a 6G communication testbed using OpenAirInterface5G and Open5GS is built on Kubernetes, as well as the system’s functionality is demonstrated via a network traffic prediction case study.
\end{abstract}


\section{Introduction}
\IEEEPARstart{I}{n} recent years, the Sixth Generation (6G) communication networks are continuously developing, with both the types and number of users growing exponentially. With billions of users and operators, network topologies are becoming increasingly complex\cite{SciChinaYou2021}. To address these problems, 6G networks must possess intelligent reasoning and self-regulation capabilities, rather than relying on manual management and preset rules. Inspired by the significant success of Artificial Intelligence (AI) in various fields, both academia and industry have begun to explore the core technologies to integrate AI into 6G networks. One of the key objectives is to enable autonomous learning and self decision-making \cite{MCHe2024}. Unlike traditional networks that simply integrate AI tools as external components, AI-native networks are designed with AI capabilities embedded at their core architecture level. This means that AI models are not merely consumers of network data but are integral to network operations, requiring real-time, structured data feeds for continuous learning and decision-making.

Recently, Software-Defined Networking (SDN) and Network Function Virtualization (NFV) allow network entities to be deployed on general hardware (e.g., X86 servers). To this end, the 3rd Generation Partnership Project (3GPP) introduced the Service-Based Architecture (SBA) for the Fifth Generation (5G) Core Networks (CN), where all Network Entities (NEs) are decomposed into individual Network Functions (NFs) \cite{3GPPsba}. Currently, most commercial CN deployments rely on virtualization platforms such as OpenStack, but are gradually transitioning to containerized and microservices-based solutions using platforms like Kubernetes. Softwarization is also advancing in access networks, such as in Open Radio Access Network (O-RAN), etc. These schemes eliminate the development and deployment gap between NEs and AI models introduced at the network design stage.

The provision of large amounts of high-quality, real-time data is crucial for the integration of AI and 6G networks \cite{WNHe2024}. To this end, 3GPP has introduced the Network Data Analytics Function (NWDAF), which enhances the intelligence of 6G networks by collecting data from various sources, including the CN, RAN, and Operation Administration and Maintenance (OAM) interfaces. However, there is no widely accepted solution for designing and deploying data collection methods in AI-native networks. Current research monitors networks and collects data by capturing packets exchanged between NEs \cite{GlobcomMan2022}, \cite{ISNCCQuad2023}, typically using Deep Packet Inspection (DPI). Although the solution is effective, DPI not only consumes additional resources and time but also operates asynchronously with NE operations. To address these limitations, this paper proposes a parallel architecture for real-time data collection during signal processing of AI-native networks. The specific contributions are summarized as:
\begin{itemize}
  \item A parallel architecture of signal processing and real-time data collection is proposed, enabling continuous data acquisition during the bitstream parsing process without additional resource and time consumption for 6G AI-native networks.
  \item A testbed is implemented on a Kubernetes platform using OpenAirInterface5G (OAI) and Open5GS, demonstrating the feasibility of the proposed parallel architecture in real-world environments.
\end{itemize}

\section{Related Work}
Currently, numerous studies have proposed solutions for building flexible, scalable, and more comprehensive 6G communication testbeds using containerization and microservices. D. Scotece et al. \cite{ComMagSco2023} utilized the powerful container orchestration capabilities of Kubernetes to propose a low-cost 5G CN deployment solution. A. Khichane et al. \cite{NOMSkhi2022} introduced a cloud NF orchestrator framework based on enhanced Kubernetes, which can effectively coordinate the network resource configurations. N. Apostolakis et al. \cite{ComMagApo2022} developed the 5G network channel simulation function and a self-developed GNU radio companion broker module through open-source software, addressing the gap of lacking dedicated hardware. W. Lee et al. \cite{ICACTLee2019} proposed a network slice architecture in 5G communication system based on the open network automation platform. These studies have made valuable contributions to enabling rapid, convenient, and low cost deployment of cloud native 5G networks. However, they do not address the need of comprehensive, systematic monitoring and management of 5G networks.

On the other hand, S. Barrachina-Muoz et al. \cite{CSNDSPBar2022} designed a cloud native 5G framework with end-to-end monitoring and over-the-air transmission functions through containerized Kubernetes cluster deployment. O. Ungureanu et al. \cite{commUng2022} proposed a new strategy for the design of 5G NextGen core functions based on cloud native approach, achieving real-time monitoring of traffic distribution, request duration and throughput. M. Mekki et al. \cite{TNSMMek2022} established a scalable network slice monitoring framework suitable for mobile networks by deploying a dedicated data collector for each network slice through Prometheus. Although existing studies provide flexible deployment and management capabilities, they lack integrating the real-time data collection with intelligent decision-making and AI model training. In-band Network Telemetry (INT) is currently widely used for network state monitoring. P. Janakaraj et al. \cite{infJan2020} implemented a distributed INT system and wireless network operating system to support machine learning algorithms in self-driving wireless networks. Z. Wang et al. \cite{tmcWang2024} proposes a network-wide data collection scheme for digital twin networks based on INT. However, INT relies on dedicated hardware and customized packet headers, hindering the usage in 6G networks with the stringent low-latency demands for large-scale data processing.

\section{Proposed Method}


\begin{figure}[t]
    \centering
    \subfloat[Data collection in traditional networks.]{
        \includegraphics[width=1\columnwidth,keepaspectratio]{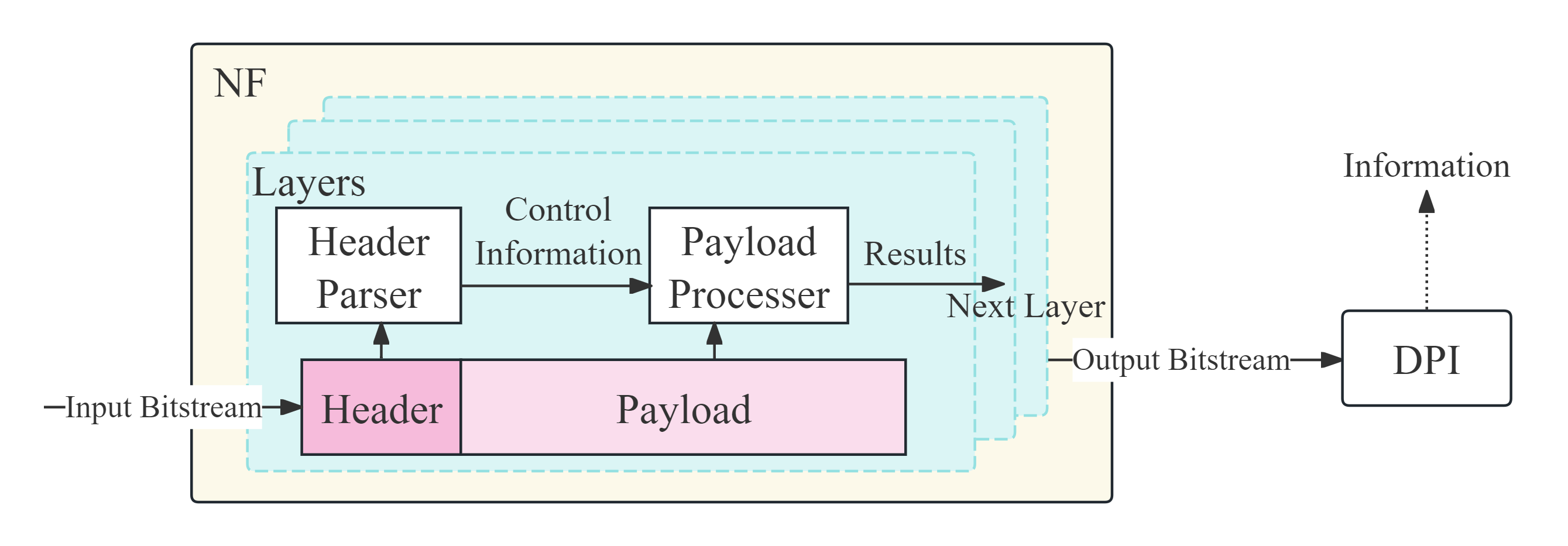}
        \label{old}
    }
    \vspace{1mm}
    \subfloat[Proposed parallel architecture.]{
        \includegraphics[width=1\columnwidth,keepaspectratio]{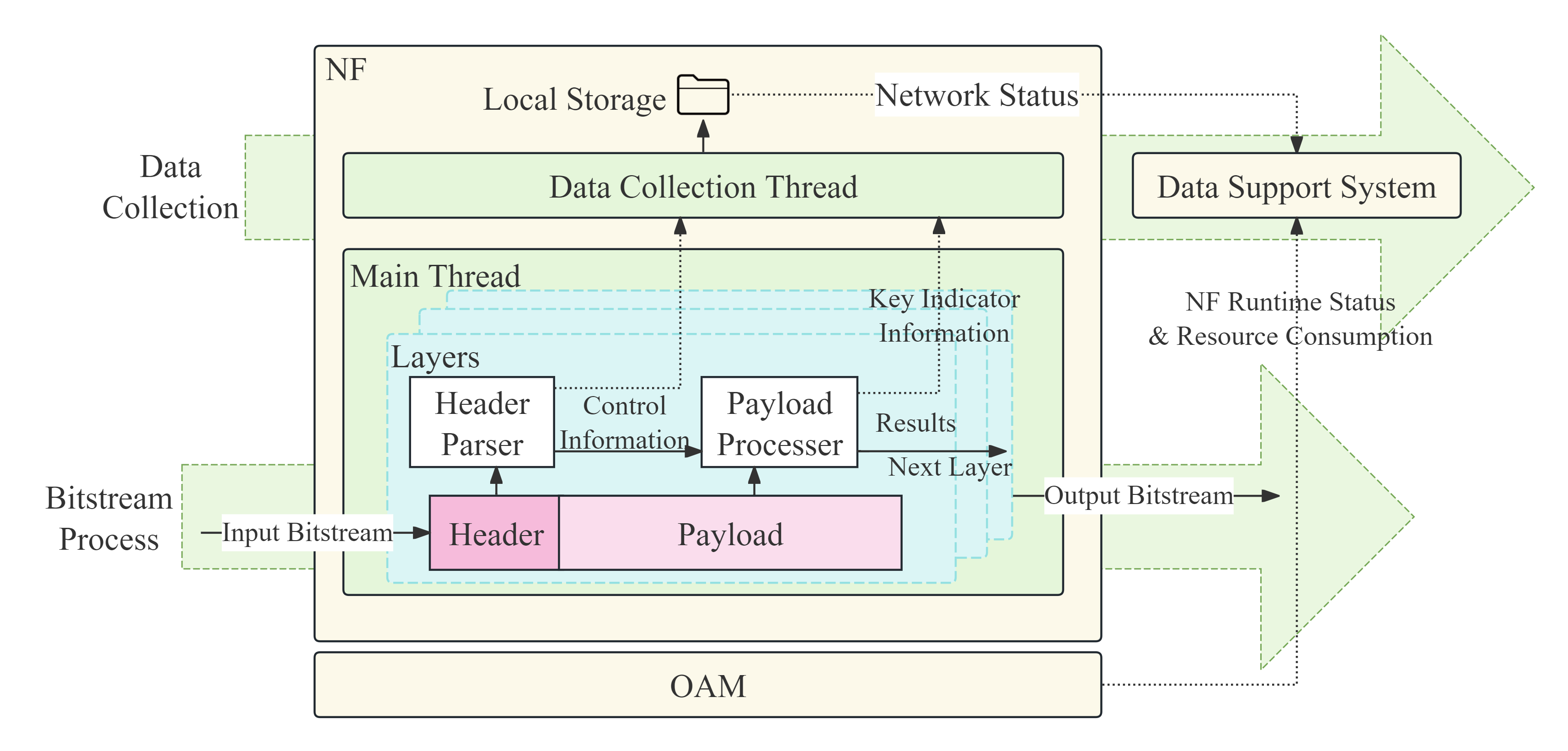}
        \label{new}
    }
    \caption{Comparison of data collection methods.}
    \label{overview}
\end{figure}

To enable data collection in real networks, this paper presents a 6G AI-native network testbed, serving as the foundation for design and implementation of data collection methods. In this testbed, NEs are implemented through containerization, with data collection capabilities integrated into the system components. The testbed is organized into three layers: the infrastructure layer, the OAM layer and the NE layer. The infrastructure layer is based on X86 hardware and Linux; OAM layer is based on Docker and Kubernetes; and the NE layer includes different NE containers based on OAI and Open5GS, with associated AI models. The infrastructure layer virtualizes hardware resources for the upper layers, while the OAM layer provides runtime environments for containers supported by virtualized resources, and manages the container lifecycle. The NEs are softwarized and deployed in containers, which enables both monitoring of each processing step within NEs and flexible management of NE containers.

In traditional networks, when a NE receives bitstreams, it parses the headers to extract control information, processes the payload based on this information, and then discards the control data and processing results once operations are complete, as shown in Fig. \ref{overview}\subref{old}. To collect this discarded data, external DPI tools such as Wireshark must be used to parse the bitstream outside the NE. This indicates that data collection is not synchronized with bitstream processing and introduces additional parsing overhead. Motivated by this observation, this study proposes a real-time data collection method with a parallel architecture, which enables data collection simultaneously with bitstream processing, as shown in Fig. \ref{overview}\subref{new}. Control information and Key Performance Indicators (KPIs) information are recorded in real-time, achieving efficient data capture without incurring extra resource consumption.

The collected data includes both network status data and system status data. The network status data is collected from different components of the network, reflecting current network operations and communication conditions, as well as the protocol stack parameters, etc. It is further categorized into CN and RAN data according to the data source. System status data refers to the status of NE entities from OAM components, including container resource consumption and container runtime status, etc. Furthermore, to automate AI model training and decision-making, a data support system based on Prometheus is developed, which can aggregate multi-level data according to the requirements of AI models and provide data retrieval services.

\begin{figure}[t]
    \centering
    \includegraphics[width=1\columnwidth,keepaspectratio]{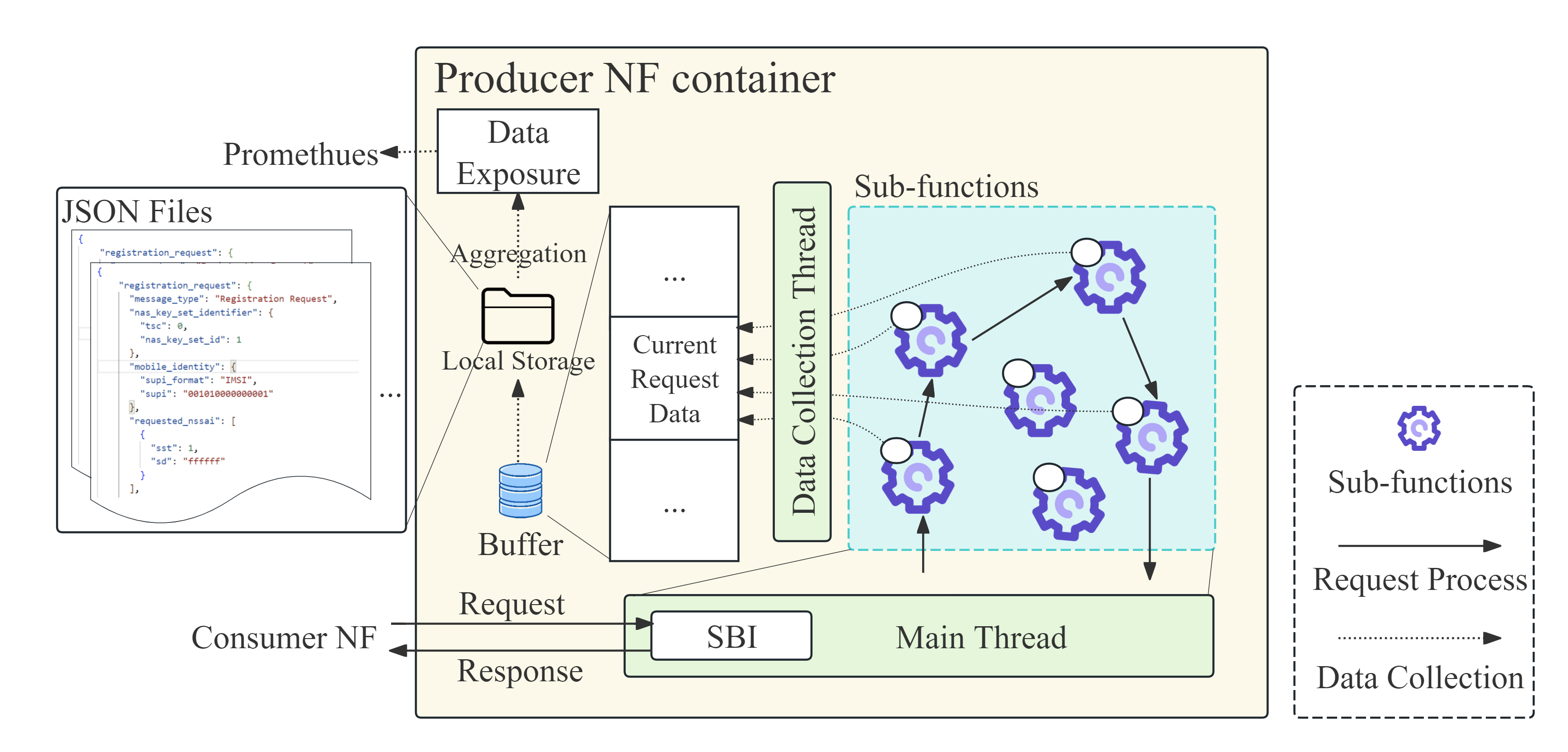}
    \caption{Data collection in CN.}
    \label{corenetwork}
\end{figure}
\subsection{Network Status Data Collection}
\subsubsection{CN data collection}
Communication between different NFs occurs through a Service Based Interface (SBI), where NFs provide services as producers and other NFs act as consumers by invoking these services as needed. Each NF container contains several sub-functions, such as, processing registration requests is a sub-function of the Access and Mobility Management Function, i.e., AMF, as shown in Fig. \ref{corenetwork}. Once an NF container starts, the main NF thread listens on the SBI for incoming requests, then sequentially invokes the relevant sub-functions according to request content and protocol requirements. Once all necessary sub-functions are executed, a response will be returned to the consumer. The data is written to the buffer as it is generated. Meantime, independent data collection threads read and store data from the buffer. Data from NF container is stored in a JavaScript Object Notation (JSON) file in local storage. There are several JSON arrays in the file, and each object contains data from one request. Additionally, a data exposure interface for Prometheus is developed, which performs statistical analysis on the raw data based on the required values. Algorithm \ref{alg:1} presents the pseudocode for real-time data collection during a service request.
\begin{algorithm}
    \caption{Data Collection in CN}
    \label{alg:1}
    \renewcommand{\algorithmicrequire}{\textbf{Input:}}
    \renewcommand{\algorithmicensure}{\textbf{Output:}}
    \begin{algorithmic}[1]
        \REQUIRE request from consumer NF 
        \ENSURE response, data of current request    
        \STATE \text{Initialize} Buffer $buf$
        \STATE \textbf{Thread 1: Main}
        \STATE $requestInfo \Leftarrow  $ ParseRequest($request$)
        \STATE $buf \Leftarrow  $ CreateMetaData($requestInfo, timestamp$)
        \STATE $subfunctionList \Leftarrow  $ MatchSubfunction($requestInfo$)

        \FOR{each $subfunction \in subfunctionList$}
            \STATE $results \Leftarrow  $ ProcessSubfunction($requestInfo$)
            \STATE $response \Leftarrow  $ UpdateResponse($results$)
            \STATE $buf \Leftarrow  $ AppendData($results$)
        \ENDFOR
        \RETURN response

        \STATE \textbf{Thread 2: Data Collection}
        \WHILE{Current request is processing}
            \STATE $data \Leftarrow  $ ReadNewData($buf$)
            \STATE $JSONArray \Leftarrow  $ ParseDataToJSON($buf$)
            \STATE $file \Leftarrow$ WriteDataToFile($JSONArray$)
        \ENDWHILE

    \end{algorithmic}
\end{algorithm}

\subsubsection{RAN data collection}
\begin{figure}[t]
    \centering
    \includegraphics[width=1\columnwidth,keepaspectratio]{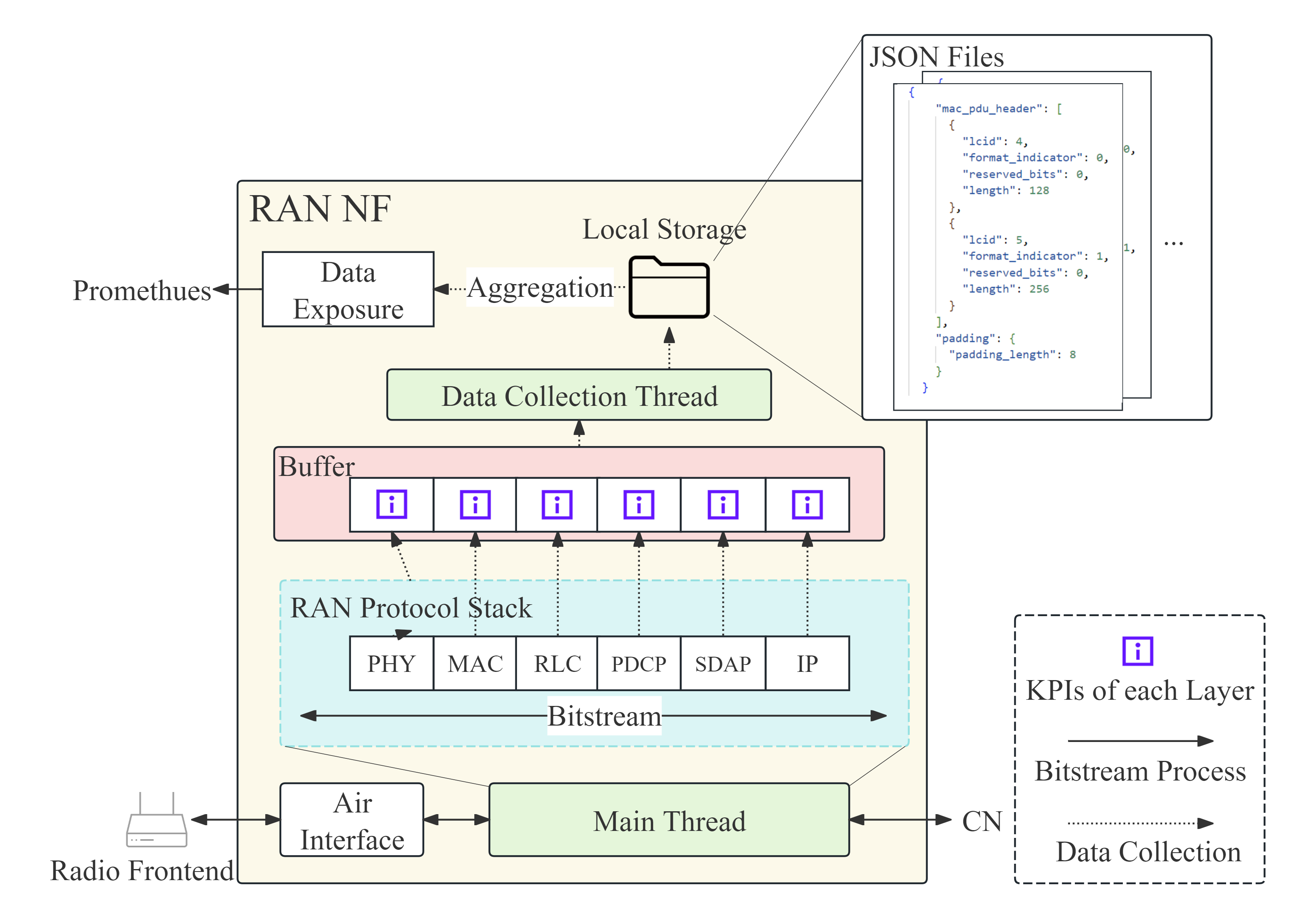}
    \caption{Data collection in RAN.}
    \label{ran}
\end{figure}

In RAN, the NEs can be categorized into Base Stations (BSs) and User Equipments (UEs). Bitstreams are continuously transmitted and received while NE is operating. Both uplink and downlink bitstreams are processed in real-time through each protocol stack. The bitstream processing adopts a parallel architecture, as shown in Fig. \ref{ran}. The main thread handles bitstream processing, while a dedicated data collection thread captures and processes the data generated at each layer, then storing it in a JSON file. Algorithm \ref{alg:2} outlines the RAN data collection workflow.

\begin{algorithm}
    \caption{Data Collection in RAN}
    \label{alg:2}
    \renewcommand{\algorithmicrequire}{\textbf{Input:}}
    \renewcommand{\algorithmicensure}{\textbf{Output:}}
    \begin{algorithmic}[1]
        \REQUIRE uplink/downlink bitstreams 
        \ENSURE processed bitstreams, data of each layer    
        \STATE \textbf{Initialize} Buffer $buf$

        \STATE \textbf{Thread 1: Main}
        \WHILE{bitstream incoming}
            \STATE $bitstream  \Leftarrow$ ReceiveBitstream()
            \FOR{each $layer \in RAN protocol stack$}
                \STATE $header, payload \Leftarrow  $ ParseHeader($layer, bitstream$)
                \STATE $results \Leftarrow  $ ProcessBitstream($header, bitstream$)
                \STATE $buf \Leftarrow  $ AppendData($header, results, timestamp$)
            \ENDFOR
            \STATE Send processed bitstream to destination
        \ENDWHILE
        \STATE \textbf{Thread 2: Data Collection}
        \WHILE{bitstream incoming}
            \FOR{each $layer \in RAN protocol stack$}
                \STATE $data \Leftarrow  $ ReadNewData($layer, buf$)
                \STATE $JSONArray \Leftarrow  $ ParseDataToJSON($data$)
                \STATE $file \Leftarrow  $ WriteDataToFile($layer, JSONArray$)
            \ENDFOR
        \ENDWHILE

    \end{algorithmic}
\end{algorithm}

The main thread processes the bitstream layer-by-layer and parses the information in the packet header. Based on the parsed content, it determines the operation and transmits this information to the data collection thread via a shared buffer. During the bitstream processing, data collection thread can receive all metadata for non-real-time functions, while for real-time functions, since there is no need to record the results of each step of bitstream processing, only key data related to signal processing is collected to reduce overhead. The data collection thread collects data from different layers separately and stores them in different files, where each JSON array records all the data involved in a single processing at the corresponding layer. Additionally, a data exposure interface is provided to allow Prometheus to access and monitor this information. It should be noted that this method can collect data from any link in the communication system, which includes not only statistical data characterizing network status but also the transmitted information itself. It can flexibly define which data to collect, and if needed, can obtain data from any link for use in AI models.

\subsection{System Status Data Collection}
\begin{figure}[t]
    \centering
    \includegraphics[width=1\columnwidth,keepaspectratio]{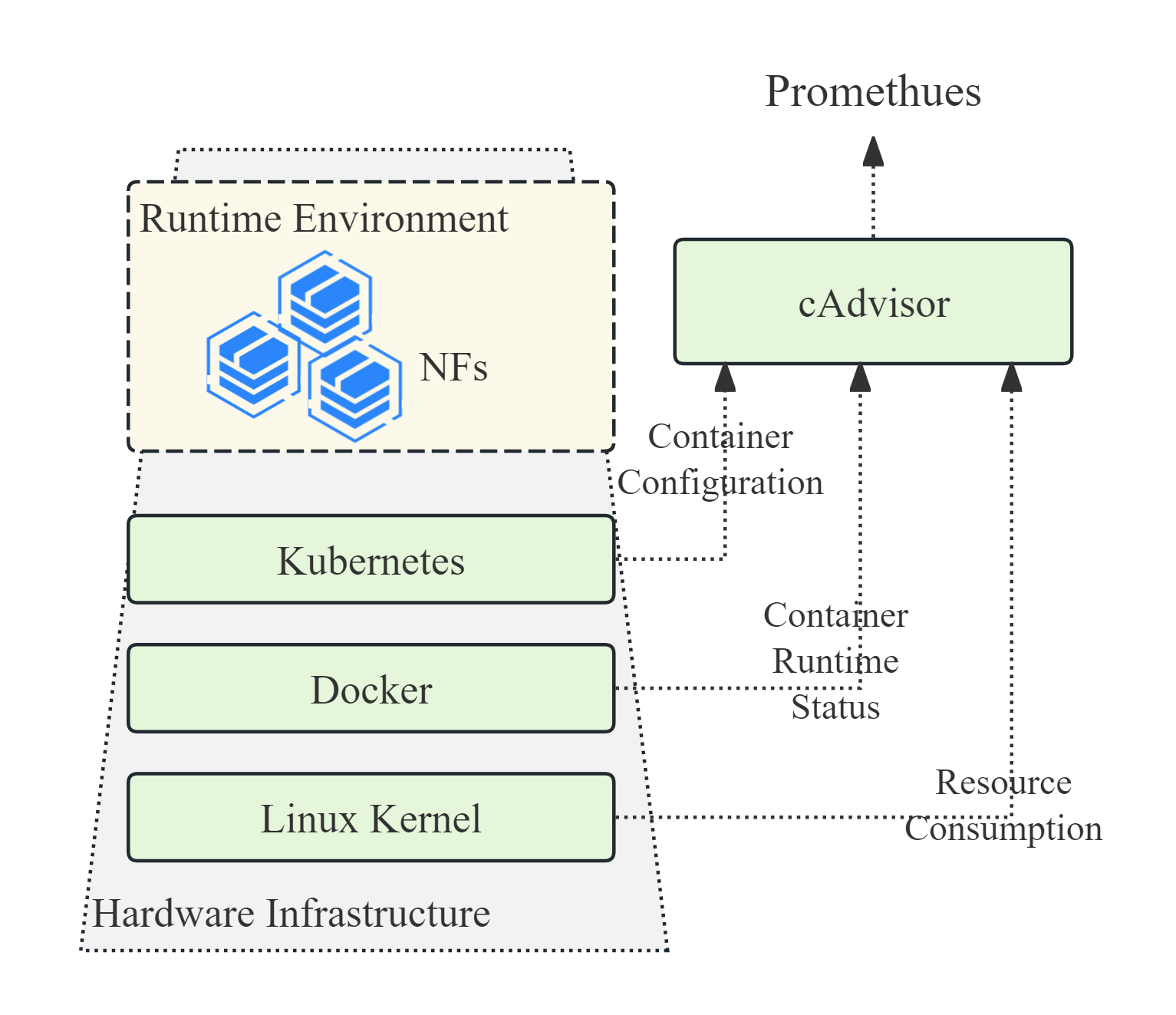}
    \caption{System status data collection.}
    \label{system}
\end{figure}

Docker and Kubernetes are used as OAM components to provide runtime environment, along with resource allocation and management for NE containers, as shown in Fig. \ref{system}. The container management relies primarily on two Linux kernel features: cgroups and namespaces. Specifically, cgroups create directory structures in file system, which contains file resource usage, such as CPU, memory, and network usage, etc. The Docker daemon obtains container resource usage information by reading these files. Meanwhile, namespaces provide an isolated execution environment, making each container appear as an independent system. Container process information can be obtained through file system, which includes detailed status information, such as process ID, status, parent process ID, etc. Container configuration and management information can be obtained via the Kubernetes APIs. The cAdvisor interacts directly with cgroups to obtain container resource usage data and retrieves container status through the Docker and Kubernetes APIs.
\subsection{Data Support for AI Model}\label{sec:support}
\begin{figure}[t]
    \centering
    \includegraphics[width=1\columnwidth,keepaspectratio]{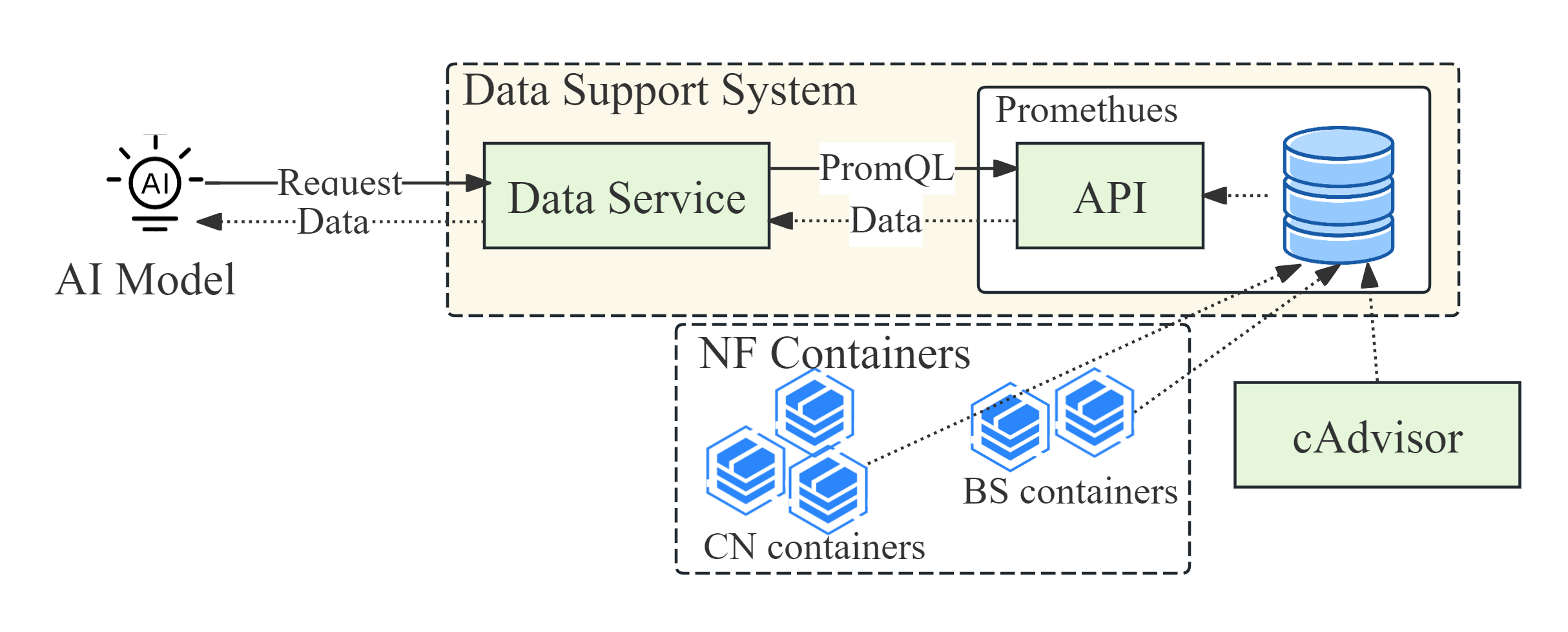}
    \caption{Data support for AI model.}
    \label{support}
\end{figure}

The data in NE's local storage contains a large amount of communication process records, which are not required by all AI models. Furthermore, this data is distributed across various NFs, posing challenges to the effective training and prediction capabilities of automated AI models. To address this issue, this paper proposes a data aggregation and storage framework based on Prometheus, and implements data service capabilities for AI models, as shown in Fig. \ref{support}. Prometheus periodically collects data from data sources including NE's data exposure interfaces and cAdvisor. The data exposure interface returns the statistical values of the preset KPIs. The raw data in the local storage of the container is transformed into queryable time series data, with the total sampling time \( T\) is divided into $n$ fixed duration time intervals $\Delta t$. Within each interval $i$, different indicators are recorded to form a sample $X_i$, denoted as $X_i = \{x_{i1}, x_{i2}, \ldots, x_{im}\}$. Additionally, a data service interface is provided for AI models, which is supported by Prometheus Query Language (PromQL). AI models retrieve data via API, while the data service function parses and manages the corresponding data access requests.

\section{System Evaluation}
To validate the effectiveness of the proposed data collection method, a testbed based on Kubernetes is established, with all functions implemented in software and running on X86 servers. Open5GS is used to implement CN functions, while OAI provides implementations for the BS and UE. An AI model is deployed on the CN node, to support intelligent capabilities within the AI-native networks. This model, based on a pre-trained TabNet regression model \cite{aaai2021}, predicts end-to-end traffic by analyzing the current state of the testbed provided by the data support system. The inputs of the AI model are formatted in a tabular structure, as illustrated in Section \ref{sec:support}.

\begin{table*}[htbp]
\caption{Overview of collected data.}
\begin{center}
\begin{tabular}{|c|c|c|c|c|c|}
\hline
\multicolumn{2}{|c|}{\textbf{Category}}&\textbf{Source}&\textbf{Function}&\textbf{Number}&\textbf{Example} \\
\hline
\multirow{16}{*}{\makecell{Network \\ Status \\ Data}} & \multirow{9}{*}{CN Data} & \multicolumn{1}{c|}{\multirow{4}{*}{AMF}} & Registration Management & 10 & amf\_reg\_reginitsucc \\ \cline{4-6}
                                                                       & & & Connection Management & 4 & amf\_ue\_connected \\ \cline{4-6}
                                                                       & & & Reachability Management & 4 & amf\_reach\_pagingsucc \\ \cline{4-6}
                                                                      &  & & Mobility Management & 6 & amf\_moblie\_ueconfigupdatesucc \\ \cline{3-6}
                   & & \multicolumn{1}{c|}{\multirow{3}{*}{SMF}}  & PDU Session Management & 22 & smf\_pdu\_pdusession\_creationsucc \\ \cline{4-6}
                                                                  &  & & GTP-U Tunnel Management & 4 & smf\_gtp\_tunnel\_active \\ \cline{4-6}
                                                                  &  & & Downlink Notification Management & 2 & smf\_dl\_manage\_rrcidle\_ues \\ \cline{3-6}
                  &  & \multicolumn{1}{c|}{\multirow{2}{*}{UPF}}  & User Plane Management & 5 & upf\_n4\_sessionreportsucc \\ \cline{4-6}
                                                                  &  & & Packet Management & 10 & upf\_n3\_gtp\_indatapakets \\ \cline{2-6}
 & \multirow{7}{*}{RAN Data} & \multicolumn{1}{c|}{\multirow{3}{*}{Data Link Layer}} & PDCP & 16 & pdcp\_rxpdu\_reorder\_count \\ \cline{4-6}
                                                                                &    & & RLC & 31 & rlc\_txpdu\_transmit\_bytes \\ \cline{4-6}
                                                                                 &   & & MAC & 56 & mac\_sched\_pucch\_size \\ \cline{3-6}
                   &  & \multicolumn{1}{c|}{\multirow{4}{*}{Physical Layer}} & Coding & 3 & phy\_max\_ldpc\_iterations \\ \cline{4-6}
                                                                              &  & & Modulation & 3 & phy\_ofdm\_offset\_divisor \\ \cline{4-6}
                                                                               & & & Resource Mapping & 31 & phy\_max\_num\_pucch \\ \cline{4-6}
                                                                               & & & Antenna Mapping & 4 & phy\_rx\_total\_gain\_dB \\ \cline{3-6}
\hline
\multicolumn{2}{|c|}{\multirow{4}{*}{System Status Data}} & \multirow{2}{*}{Kubernetes} & Pods Deployment & 22 & kube\_pod\_container\_status\_ready \\ \cline{4-6}
                                          \multicolumn{2}{|c|}{}  & & Pods Configuration & 2 & kube\_pod\_container\_resource\_limits \\  \cline{3-6}
                 \multicolumn{2}{|c|}{}   & \multirow{1}{*}{Docker}  & Container Monitoring & 12 & container\_cpu\_usage\_seconds\_total \\ \cline{3-6}
                 \multicolumn{2}{|c|}{}   & \multirow{1}{*}{Linux Kernel}  & System Resource Monitoring & 42 & node\_network\_receive\_bytes\_total \\ \cline{3-6}
\hline
\end{tabular}
\label{tab1}
\end{center}
\end{table*}

\begin{figure}[t]
    \centering
    \includegraphics[width=1\columnwidth,keepaspectratio]{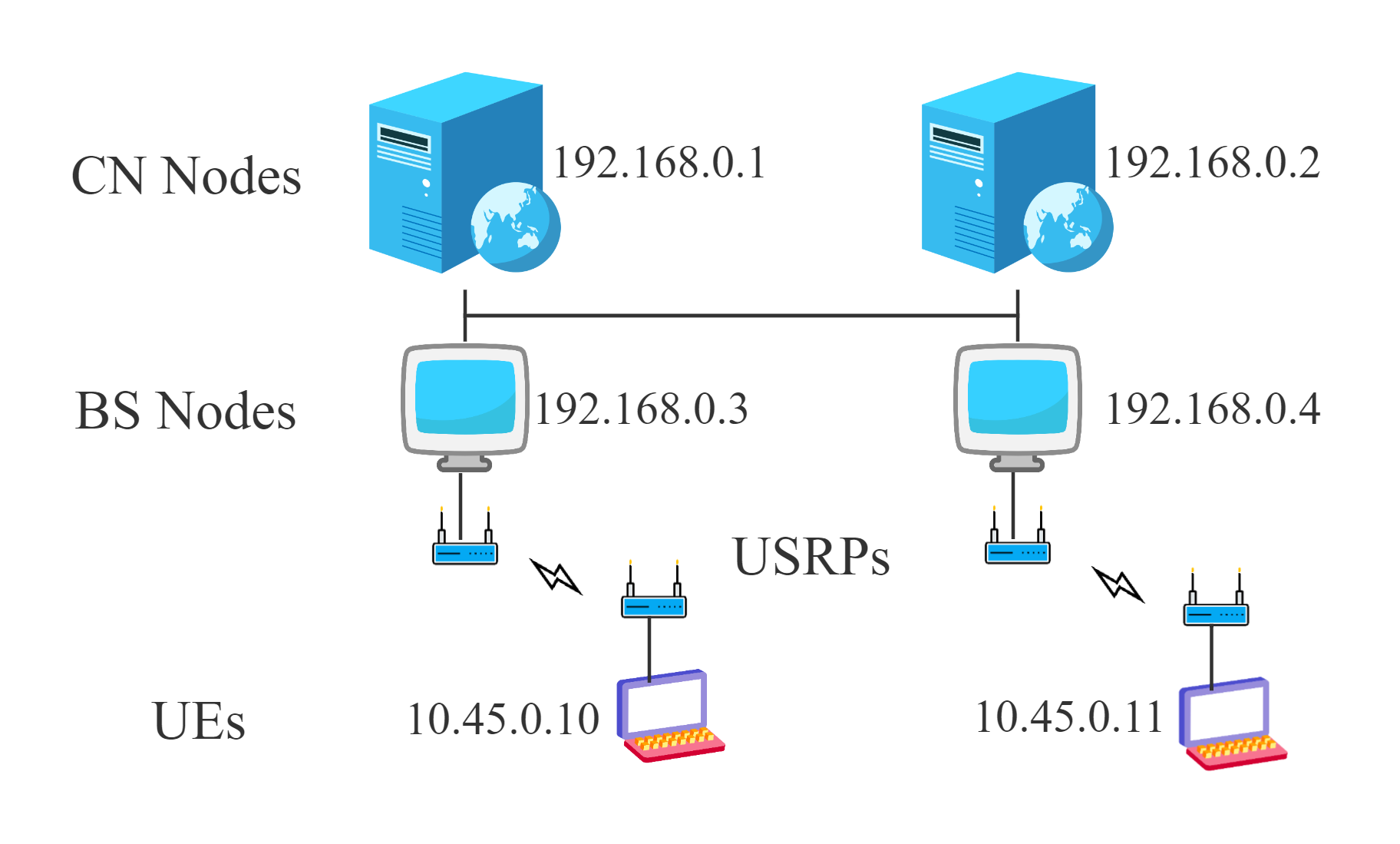}
    \caption{Testbed Structure}
    \label{testbed}
\end{figure}

The testbed includes both server-side and user-side components. Server-side functions are deployed within a Kubernetes cluster consisting of two CN nodes and two BS nodes, while the user-side components are deployed on X86-based terminals. Each BS node and UE terminal is connected to a Universal Software Radio Peripheral (USRP) B210, as shown in Fig. \ref{testbed}. After UE1 and UE2 join the Open5GS subnet, a test script based on iperf3 was deployed to generate traffic between two UEs. An iperf client was started on UE1, while an iperf server was running on UE2. The function of the script is to continuously execute iperf, establish a TCP connection, and generate traffic from UE1 to UE2. The preset bitrate sent from the client to the server is randomly determined at each execution. After starting the test script, the AI model requests data from Prometheus every minute to perform end-to-end traffic prediction.

\begin{figure*}[t]
    \centering
    \subfloat[Resource consumption of proposed method.]{
        \includegraphics[width=1.5\columnwidth,keepaspectratio]{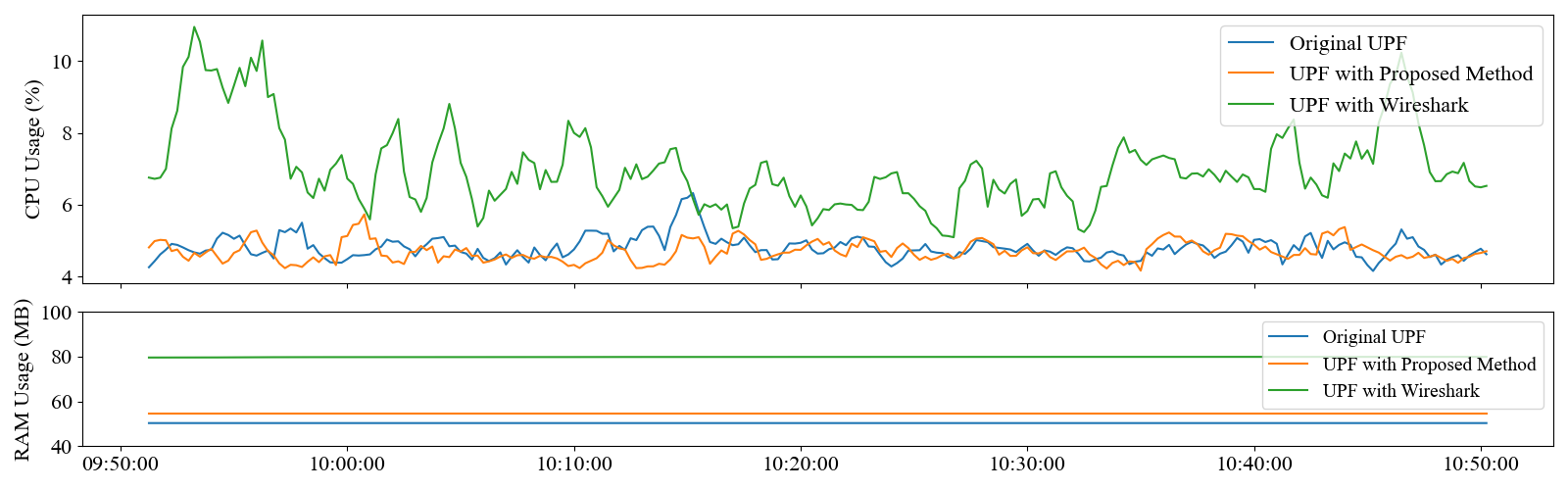}
        \label{a}
    }
    \vspace{1mm}
    \subfloat[Latency comparison.]{
        \includegraphics[width=1.5\columnwidth,keepaspectratio]{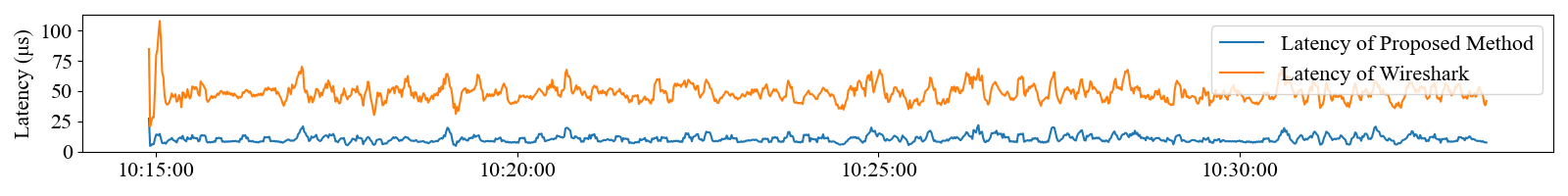}
        \label{b}
    }
    \caption{Performance evaluation.}
    \label{perform}
\end{figure*}

\begin{figure*}[t]
    \centering
    \includegraphics[width=1.5\columnwidth,keepaspectratio]{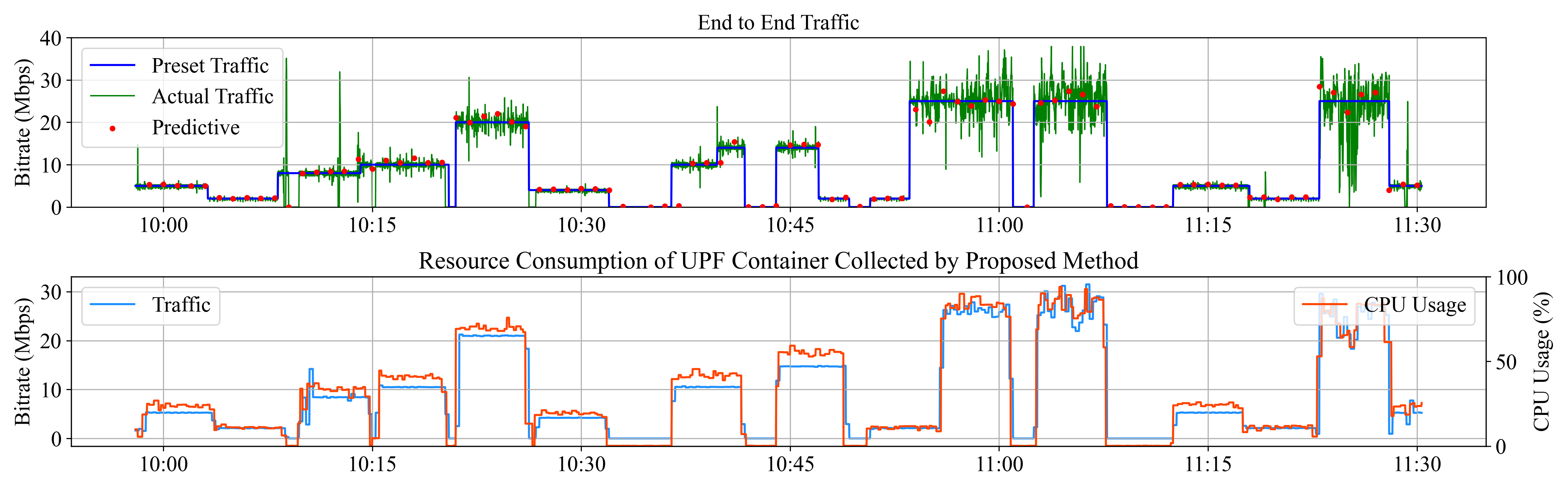}
    \caption{System level evaluation.}
    \label{system_eval}
\end{figure*}

\subsubsection{Collected Data}

During the experimental period, all network traffic was processed locally through the UEs, BSs and CN within their designated NEs, ensuring that all original communication data is stored locally. From these data, multi-level KPIs were extracted from various sources and systematically categorized according to their respective functions, as detailed in Table \ref{tab1}. This table enumerates the KPIs associated with each function and provides representative examples. All KPIs were gathered by Prometheus at a predetermined frequency, enabling AI models to access the data via Prometheus APIs.

\addtolength{\topmargin}{0.03in}
\subsubsection{Performance of Data Collection Approach}

To evaluate the performance of the proposed data collection approach, three User Plane Function (UPF) containers run simultaneously on the testbed, including an original UPF, a UPF with proposed data collection method and a UPF integrated with Wireshark. The performance of these UPF containers is shown in the Fig. \ref{perform}\subref{a}. Without significantly increasing additional resource consumption, the proposed method significantly reduced CPU resource consumption by 32.7\%, and memory resource consumption by 31.6\%. Furthermore, data collection latency was also recorded, as shown in the Fig. \ref{perform}\subref{b}. This latency is defined as the time interval from when a packet enters the UPF container until the packet's information is successfully stored in local storage. It is evident that the proposed method reduced the average delay by 78.4\% compared to Wireshark.

\subsubsection{System-level Evaluation}

After deploying the data collection method in the UPF container, the aforementioned scripts are executed. The AI model generates traffic predictions every minute based on the real-time network and system status data provided by the proposed method. TabNet identifies the correlation between each KPI and the prediction target, which is end-to-end traffic in this experiment. These correlation scores enable the selection of the most relevant KPIs for system performance evaluation, as shown in Fig. \ref{system_eval}. The preset bitrate is a randomly assigned value when the test script starts iperf, while the actual bitrate is recorded at the UE terminal. The results show that while the function of the UPF is not affected, the proposed method can support AI models in prediction and decision-making. This demonstrates that the proposed data collection method effectively reflects network status and can autonomously works in real communication networks.

\section{Conclusion}

A parallel architecture for real-time data collection in 6G AI-native networks has been designed and implemented. A comprehensive data collection system was developed to collect data from different sources and provide automatic data service for AI model training and decision-making. To validate the proposed approach, a 6G testbed was built on Kubernetes, and the system's operational process was demonstrated through a use case of network traffic prediction. Experimental results confirmed the performance and scalability of the proposed method, highlighting its potential to advance 6G AI-native network research and applications. In future work, further research will focus on enhancing data collection and validation systems, improving system scalability and elasticity, and enabling more comprehensive all-round learning.

\section{Acknowledgement}
This work was supported in part by Top-Ten Technological Projects for Hunan Province under Grant 2025QK1009 and in part by National Natural Science Foundation of China under Grant 62171474.

\bibliographystyle{IEEEtran}

\begin{thebibliography}{00}

\bibitem{SciChinaYou2021}X. You, et al., ``Towards 6G wireless communication networks: Vision, enabling technologies, and new paradigm shifts." \emph{Science China Inf. Sciences}, vol. 64, pp. 1-74, 2021.

\bibitem{MCHe2024}S. He, et al., ``Challenges and methods of constructing a verification system for endogenous intelligent communication in wireless networks," \emph{Mobile Commun.}, vol. 48, no. 7, pp. 2-14, 2024.

\bibitem{3GPPsba}3GPP TS 29.500, ``5G system: Technical realization of service based architecture; Stage 3," v1.0.0, March 2018.

\bibitem{WNHe2024}S. He, ``An endogenous intelligent architecture for wireless communication networks," \emph{Wireless Netw.}, vol. 30, Issue 2, pp. 1069–1084, 2024.

\bibitem{GlobcomMan2022}D. M. Manias, A. Chouman and A. Shami, ``An NWDAF approach to 5G core network signaling traffic: Analysis and characterization," in \emph{IEEE Global Commun. Conf.}, 2022, pp. 6001-6006.

\bibitem{ISNCCQuad2023}M. Quadrini, et al., ``Data collection using NWDAF network function in a 5G core network with real traffic," in \emph{2023 International Symposium on Netw., Computers and Commun. (ISNCC)}, 2023, pp. 1-7.

\bibitem{ComMagSco2023}D. Scotece, et al., ``5G-Kube: Complex telco core infrastructure deployment made low-cost," \emph{IEEE Commun. Mag.}, vol. 61, no. 7, pp. 26-30, 2023.

\bibitem{NOMSkhi2022}A. Khichane, et al., ``Cloud native 5G: An efficient orchestration of cloud native 5G system," in \emph{IEEE/IFIP Netw. Operations and Manag. Symp.}, 2022, pp. 1-9.

\bibitem{ComMagApo2022}N. Apostolakis, M. Gramaglia and P. Serrano, ``Design and validation of an open source cloud native mobile network," \emph{IEEE Commun. Mag.}, vol. 60, no. 11, pp. 66-72, 2022.

\bibitem{ICACTLee2019}W. Lee, T. Na and J. Kim, ``How to create a network slice? - A 5G core network perspective," in \emph{the 21st Inter. Conf. on Advanced Commun. Techn. (ICACT)}, 2019, pp. 616-619.

\bibitem{CSNDSPBar2022}S. Barrachina-Mu\~{n}oz, M. Payar\'{o} and J. Mangues-Bafalluy, ``Cloud-native 5G experimental platform with over-the-air transmissions and end-to-end monitoring,'' in \emph{the 13th Inter. Symp. on Commun. Syst., Netw. and Digital Signal Proc. (CSNDSP)}, 2022, pp. 692-697.

\bibitem{commUng2022}O. Ungureanu and C. Vl\u{a}deanu, ``Leveraging the cloud-nati ve approach for the design of 5G NextGen core functions,'' in \emph{the 14th Inter. Conf. on Commun. (COMM)}, 2022, pp. 1-7.

\bibitem{TNSMMek2022}M. Mekki, S. Arora and A. Ksentini, ``A scalable monitoring framework for network slicing in 5G and beyond mobile networks," \emph{IEEE Trans. on Netw. and Service Manag.}, vol. 19, no. 1, pp. 413-423, 2022.

\bibitem{infJan2020}P. Janakaraj, et al., ``Towards In-Band Telemetry for self driving wireless networks," in \emph{Proc. of the IEEE Conf. on Computer Commun. (INFOCOM)}, 2020, pp. 766-773.

\bibitem{tmcWang2024}Z. Wang, D. Jiang, and S. Mumtaz, ``Network-wide data collection based on in-band network telemetry for digital twin networks," \emph{IEEE Trans. on Mobile Comput.}, vol. 24, no. 1, pp. 86-101, 2025.

\bibitem{aaai2021}S. Arik and T. Pfister, ``TabNet: Attentive interpretable tabular learning,'' in \emph{Proceedings of the AAAI Conf. on Artificial Intelligence}, vol. 35, no. 8, 2021, pp. 6679-6687.


\end{thebibliography}
\balance

\end{document}